# Radio frequency polarization modulation based on an optical frequency comb

Ruixue Zhang,[1, 2] Yiming Gong,[1] Matthew W. Day,[1] Dong Sun,[1,3, a)] and Steven T. Cundiff[1, b]

1) Department of Physics, University of Michigan, Ann Arbor, MI 48109, USA

2) State Key Laboratory of Precision Measurement Technology and Instruments, Department of Precision Instrument, Tsinghua University, Beijing 100084, China

3) International Center for Quantum Materials, School of Physics, Peking University, Beijing 100871, China

a)sundong@pku.edu.cn
b)cundiff@umich.edu

**Abstract**

We propose a method to generate stabilized radio-frequency polarization modulation based on optical frequency combs. Two pulse trains with the same repetition rate and different offset frequencies generate arbitrary polarization states that are modulated at the offset frequency difference. Long-term stability of the polarization modulation is demonstrated with the modulation frequency at $f_{rep}/2$. Modulation at $f_{rep}/4$ is also demonstrated to show the flexibility of the technique. We employ an electrical delay line to fine-tune the polarization states that constitute the time-dependent modulation.

**Introduction**

Direct manipulation and periodic modulation of the polarization state produced by an ultrafast laser is desirable for a wide range of spectroscopy and imaging applications when the material interaction with light is polarization-sensitive.[1-7] This capability has recently become more important for studying quantum materials with the non-trivial spin, valley, or chiral properties that are sensitive to the light helicity.[8-12] Furthermore, materials with anisotropic polarization responses are of great interest in a variety of novel applications.[13-16] Rapid polarization modulation is required when ultralow noise or ultrafast time resolution is desired.[7] The most straightforward polarization modulation scheme is to mechanically rotate a waveplate, but this method is unstable, and the modulation frequency is low, limited by the mechanics.[17-21] Compared to mechanical modulation, polarization modulation based on photoelastic modulator is more stable and can reach higher modulation frequency, but the polarization modulation frequency is still limited to tens of kHz.[22-24]

The development of optical frequency combs (OFCs) has provided a new paradigm to reach radio frequency (RF) polarization modulation. OFCs offer a high degree of stability and control over the detailed properties of their spectra.[25,26] They typically offer hundreds of thousands of mutually coherent low-noise laser lines over a broad spectrum. The frequency of each mode is determined by the repetition frequency ($f_{rep}$) and carrier-envelope offset frequency ($f_{ceo}$) such that $f(n) = nf_{rep} + f_{ceo}$. By tightly stabilizing $f_{rep}$ and $f_{ceo}$, the light source can be treated as a precise "frequency ruler", where the absolute frequency of each longitudinal mode is accurately determined with respect to all of the other longitudinal modes.[27,28] Recently, the OFCs have been recognized as a powerful tool and applied in various applications including absolute distance ranging,[29–31] surface-profile metrology,[32–34] and spectroscopy.[35–38] In the time domain, $f_{ceo}$ corresponds to a pulse-to-pulse evolution of the carrier–envelope phase (CEP), $\Phi_{CEP,p-p} = 2\pi f_{ceo}/f_{rep}$. This feature can be utilized for coherent control of CEP dependent phenomena, such as quantum interference.[39,40] By precisely controlling the difference of $f_{ceo}$ ($\Delta f_{ceo}$) of two combs with the same $f_{rep}$, the combined electric fields the two combs can produce high-frequency polarization modulations with a modulation frequency of $\Delta f_{ceo}$, which can reach the order of ~GHz depending on the comb sources that are used.[41-44] Precise control of the relationships between $f_{rep}$, $f_{ceo}$ and $\Delta f_{ceo}$ enables the versatile design of tailored polarization modulation schemes as demonstrated in this work.

Here we demonstrate a highly stabilized 46.735-MHz polarization modulation with two polarization states that have a π-phase difference in each modulation period, e.g. pulse-to-pulse switching between left/right circular polarizations or horizontal/vertical linear polarizations. This RF polarization-modulated ultrafast light source is based on combining two orthogonally polarized combs generated by splitting a 93.47-MHz $f_{rep}$-stabilized optical frequency comb centered at 800 nm. By using two acousto-optic modulators (AOMs) to impart a controllable phase ramp between the two frequency combs, we generate a stable polarization modulation. We further demonstrate that the polarization scenario can be conveniently stabilized and tuned by locking to a stable frequency reference with a tunable phase. On the other hand, by tuning the $\Delta f_{ceo}$ of the two arms, the polarization modulation can be tuned to switch between $N$ different polarizations that are $2\pi/N$ phase apart in each modulation period with a modulation frequency of $f_{rep}/N$. Beyond $N = 2$, $N = 4$ is also experimentally demonstrated as an example. The demonstration of this highly stabilized radio frequency polarization modulation approach based on an OFC paves the way toward advanced polarization-sensitive spectroscopy and imaging instrumentation with ultralow noise and ultrafast speed.

## II. SYSTEM SETUP AND METHOD OF OPERATION

**Radio Frequency Polarization Modulation Generation**

To achieve polarization modulation, the linear-polarized optical-frequency-comb output from a homebuilt Kerr-lens mode-locked Ti: sapphire laser centered at 800 nm (~375 THz) is split into two arms as shown in Fig. 1(a). The repetition frequency ($f_{rep}$) of the comb is locked to a direct digital synthesizer at 93.47 MHz using a feedback circuit and a piezoelectric transducer to control the cavity length. The offset frequency ($f_{ceo}$) is not actively controlled. The linear polarization of one arm is rotated by a half wave plate (HWP) to be perpendicular to the polarization of the other arm, then the two orthogonally polarized comb pulses ($E_x\hat{x}$ and $E_y\hat{y}$) are recombined with a polarizing beam splitter (PBS). A delay stage is placed in one arm to control the temporal overlap of the combined pulses. The polarization scenario of the synthesized comb can be described by the Jones vector

$$E(t) = \begin{bmatrix} E_x \\ E_y e^{i[\Delta\Phi_{CEP}(t) + \Delta\Phi_l]} \end{bmatrix}, \tag{1}$$

where $\Delta\Phi_{CEP}(t) = 2\pi\Delta f_{ceo}t$ and $\Delta\Phi_l = 2\pi\Delta l/\lambda$ accounts for the phase difference due to the relative carrier-envelope phase (CEP) and optical path length difference ($\Delta l$) between the two arms respectively and $t$ denotes the time position of a pulse in the synthesized comb. The phase shift between two successive pulses of the synthesized comb is $2\pi\Delta f_{ceo}/f_{rep}$. In all cases reported below, we tune the power in each arm such that $|E_x|=|E_y|$, as plotted in Fig. 1(c). Circular polarization is obtained when $\Delta\Phi = \Delta\Phi_{CEP}(t) + \Delta\Phi_l = \pm\pi/2$. Since $\Delta\Phi$ evolves with $t$, the phase difference between the polarizations of adjacent pulses can be written as $2\pi/N$, with $N = f_{rep}/\Delta f_{ceo}$. If $N$ is an integer, which can be achieved by precise control of $\Delta f_{ceo}$, then the polarization state in Eq. (1) evolves with frequency $f_{rep}/N$.

To achieve the desired polarization modulation, an AOM is inserted in each arm of the interferometer, providing control over $\Delta f_{ceo}$. To obtain polarization modulation with $N = 2$, the two combs were diffracted into +1st and -1st order of the two AOMs respectively with $f_{AOM1} = f_{AOM2} = $ 70.1025 MHz, where $f_{AOM1}$ and $f_{AOM2}$ are the driving frequencies of the two AOMs. Therefore, $\Delta f_{ceo}$ = mod($f_{AOM1}+f_{AOM2}$, $f_{rep}$) = 46.735MHz = $f_{rep}/2$, yielding a $\Delta\Phi_{CEP}$ of $\pi$ for the two adjacent synthesized pulses.

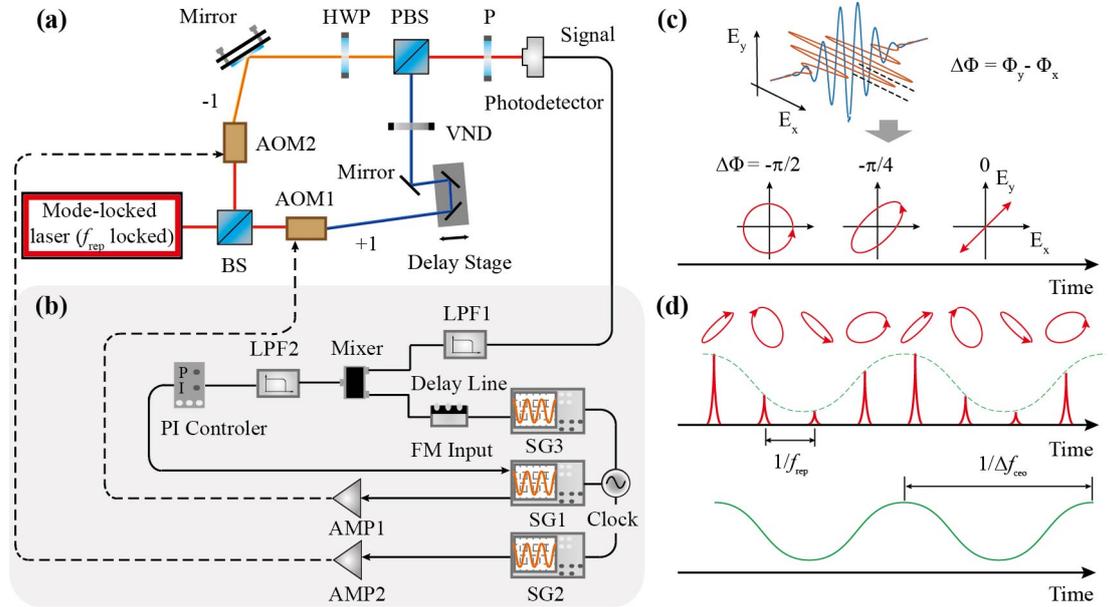

FIG. 1. Experimental configuration and the principle of polarization modulation. (a) Schematic diagram of the optical setup to obtain polarization modulation. BS: beam splitter. HWP: half wave plate. VND: variable neutral density filter. (b) Electrical feedback loop used to stabilize the phase evolution of the polarization modulation period. AMP: RF power amplifier. (c) The operational principle of synthesizing a polarization-modulated comb from two orthogonally polarized combs, $E_x\hat{x}$ and $E_y\hat{y}$, with the phase difference of $\Delta\Phi$. Different polarization states can be generated depending on the value of $\Delta\Phi$. (d) Schematic diagram of the polarization modulated combs ($N = 4$) sampled by a 45° linear polarizer (P) followed by a fast PD (the upper panel) and filtered signal after LPF1 (the bottom panel).

**Polarization Stabilization and Control**

While the change in relative phase difference of the orthogonally polarized combs for successive pulses is $\Delta\Phi_{CEP} = 2\pi/N$, the polarization scenario, that is whether it switches between left/right circular polarizations or two perpendicular linear polarizations when $N = 2$, is determined by $\Delta\Phi_l$, which fluctuates with the relative optical path length difference ($\Delta l$) of the two arms. Additional phase fluctuations could also come from the frequency fluctuation of $f_{ceo}$ due to noise in the AOM modulation frequency. Unless $\Delta\Phi$ is stabilized, the polarization scenario evolution of the synthesized comb fluctuates. To stabilize $\Delta\Phi$, the drift of $\Delta\Phi$ must be monitored and an error signal must be generated and fed back to either the optical path difference ($\Delta\Phi_l$ term) or $\Delta f_{ceo}$ ($\Delta\Phi_{CEP}$ term) to compensate the phase drift. In this work, we choose to feedback to the $\Delta\Phi_{CEP}$ term because the bandwidth of a mechanical feedback loop would be orders of magnitude lower than that of a purely electronic feedback loop.

To monitor the drift of $\Delta\Phi$ ($\delta\Delta\Phi$), the synthesized comb is sampled with a polarizer (P) whose axis is set 45° relative to both orthogonally polarized combs followed by a fast photo detector (PD) with bandwidth much larger than $f_{rep}$. In this way, the polarization modulation of the synthesized comb is converted to RF voltage modulation in the PD output (upper panel of Fig. 1(d)). Then the PD output is the input to an electrical feedback loop as shown in Fig. 1(b). In the feedback loop, three radio-frequency signal generators (SG) use a common 10 MHz Rb frequency standard. Two SGs, both set at 70.1025 MHz, are amplified to drive the AOMs and the third SG generates a phase reference signal at the frequency of $\Delta f_{ceo}$ (46.735MHz): $R(\Delta f_{ceo}, \Phi_R) \sim \sin(2\pi\Delta f_{ceo}t + \Phi_R)$, where $\Phi_R$ is the reference phase that is tunable by an electric delay line after SG3. To generate the error signal, the PD output goes through an electronic low pass filter (LPF1) that blocks the $f_{rep}$ component but passes $\Delta f_{ceo} = f_{rep}/N$. The output signal after LPF1 provides a signal at the frequency of $\Delta f_{ceo}$: $E(\Delta f_{ceo}, \delta\Delta\Phi) \sim \sin(2\pi\Delta f_{ceo}t + \Delta\Phi_l + \delta\Delta\Phi)$ which carries polarization fluctuation signal ($\delta\Delta\Phi$) in its phase term (lower panel of Fig. 1(d)). Then $E(\Delta f_{ceo}, \delta\Delta\Phi)$ is mixed with $R(\Delta f_{ceo}, \Phi_R)$ and the mixer output passes another low pass filter (LPF2) that only passes the DC component. This DC component after LPF2 provides the error voltage signal: $V_{error} \sim \cos(\Delta\Phi_l + \delta\Delta\Phi - \Phi_R)$. $V_{error}$ is amplified and integrated by a proportional-integral (PI) controller, the output of the PI controller controls the frequency modulation (FM) input of SG1, which converts $V_{error}$ into a frequency shift $\delta f \sim V_{error}$ and adjusts the offset frequency $\delta f$ to AOM1. The modification of the $f_{ceo}$ of one comb by $\delta f$ compensates the overall phase drift of $\Delta\Phi(t)$. When the feedback loop is engaged, $V_{error}$ is locked to zero, so the $\Delta\Phi_l + \delta\Delta\Phi - \Phi_R$ term is locked to $\pi/2$. To tune the polarization scenarios of the synthesized comb ($\Delta\Phi$), one can conveniently tune $\Phi_R$ by the electrical delay line after SG3. In this way, a highly stabilized, continuously tunable polarization-modulated comb is generated.

### III. EXPERIMENTAL RESULTS
#### A. Characterization of RF polarization modulated comb
To characterize and monitor the synthetized polarization modulated comb, the polarization modulation is projected on the 45° linear polarization direction by a polarizer to be converted to amplitude modulation. The output from a fast photo detector (PD) is digitized by a data acquisition board (DAQ) at a sampling rate of 1.5 GS/s. The Jones vector of the polarization states of the synthesized pulses, the polarization scenario, of the synthesized comb is

$$E(t) = \begin{bmatrix} E_x \\ E_y \end{bmatrix} = a \begin{bmatrix} 1 \\ \exp[i(2\pi\Delta f_{ceo}t + \Delta\Phi_l + \delta\Delta\Phi)] \end{bmatrix}, \qquad (2)$$

where $a^2$ is the ratio of the power present in each arm of the interferometer. After passing through a linear polarizer inclined at angle $\theta$, the field detected by the PD is

$$\begin{bmatrix} E_{out}(t) \\ 0 \end{bmatrix} = \mathbf{P} \times \mathbf{R}(\theta) \times \begin{bmatrix} E_x \\ E_y \end{bmatrix} = \begin{bmatrix} 1 & 0 \\ 0 & 0 \end{bmatrix} \times \begin{bmatrix} \cos(\theta) & \sin(\theta) \\ -\sin(\theta) & \cos(\theta) \end{bmatrix} \times$$
$$a \begin{bmatrix} 1 \\ \exp[i(2\pi \Delta f_{ceo} t + \Delta \Phi_l + \delta \Delta \Phi)] \end{bmatrix} = \begin{bmatrix} \sqrt{2} \times a \times [1 + \exp[i(2\pi \Delta f_{ceo} t + \Delta \Phi_l + \delta \Delta \Phi)]] \\ 0 \end{bmatrix}. \quad (3)$$

With the angle of the polarizer set to $\theta = 45°$, and the power of each arm of the interferometer set to be equal ($a = 1$), the output intensity detected by the reference PD is

$$I_{out}(t) = 1 + \cos(2\pi \Delta f_{ceo} t + \Delta \Phi_l + \delta \Delta \Phi). \quad (4)$$

When the feedback loop is engaged, the sum of $\Delta \Phi_l$ and $\delta \Delta \Phi$ is locked to a fixed value $\pi/2 + \Phi_R$, where $\Phi_R$ represents the phase set by the reference signal generator (SG3 in Fig. 1). Therefore, the polarization modulation of the synthesized comb is converted to the amplitude modulation over time with period of $1/\Delta f_{ceo}$.

Taking $N = f_{rep}/\Delta f_{ceo} = 2$ as an example, there are $N = 2$ synthesized pulses in each modulation period, and the phase of the two successive polarization states shifts $2\pi/N = \pi$. The intensities of the two adjacent pulses acquired by the PD are

$$I_{out1} = 1 + \cos(\Delta \Phi\ ), \quad (5)$$

and

$$I_{out2} = 1 - \cos(\Delta \Phi\ ). \quad (6)$$

The amplitude ratio ($R$) of two adjacent peaks in a single modulation period is

$$R = [1 - \cos(\Delta \Phi\ )]/[1 + \cos(\Delta \Phi\ )]\ . \quad (7)$$

Experimentally, the amplitude ratio $R$ can be obtained directly from the amplitudes of the two adjacent pulses acquired by the DAQ. However, direct extraction of $\Delta \Phi$ from Eq. (7) is not feasible and reliable because a nonzero noise floor exists in the PD response of all pulses as shown in Fig. 2. The noise floor never vanishes due to combined artifacts in the response of PD and DAQ, which leads to a systematic error in determining $\Delta \Phi$ through Eq. (7). Even so, the fluctuation of $\Delta \Phi$ ($\delta \Delta \Phi$), which propagates into the fluctuation of $R$ ($\delta R$), still reflect the stability of the generated polarization states, and $\delta \Delta \Phi$ is connected to $\delta R$ through

$$\delta \Delta \Phi = \delta R / [\sqrt{R}(1 + R)]\ . \quad (8)$$

**B. Stabilized polarization modulation with $N = 2$**

Characterization of a phase-stabilized polarization modulated pulse train with $N = 2$ is presented in Fig. 2. Six sets of data with time span of 20 ms spread over 3 minutes are recorded. Figure 2(a) shows amplitude modulation at the frequency of $f_{rep}/2$ for 120-ns time frames within each 20-ms record. The two peaks within one modulation period represent different polarization states with a π-phase shift. Figure 2(b) shows the average and standard deviation of the amplitudes of the two polarization peaks for each 20-ms time span. The ratio ($R$) between the two peaks, which characterizes the polarization modulation scenario as defined by Eq. (7), slightly fluctuates within each 20 ms, but it kept around a constant over the whole 3 minutes as shown in Fig. 2(c), as a result of the electric feedback loop. The fluctuation of $\Delta\Phi$ ($\delta\Delta\Phi$) can be obtained from the standard deviation of $R$ through Eq. (8), which is less than 2.6°. These results indicate that the polarization evolution is highly stabilized during the 3-minutes measurement period. We observe such highly stable polarization modulation lasting over hours.

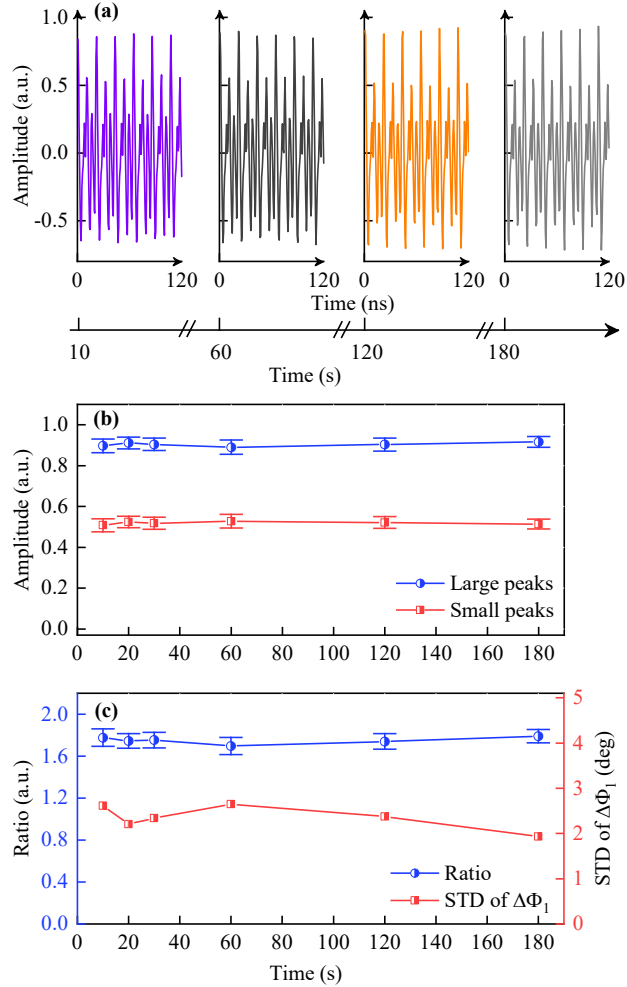

FIG. 2 Long-term stability of the polarization state modulation. (a) 120-ns snapshots of the projected pulse amplitude sequence at around 10 s, 60 s, 120 s, and 180 s. Polarization modulation at the frequency of $f_{rep}/2$ can be observed. (b) Amplitude fluctuation of the large pulse peaks (blue) and small pulse peaks (red) for each set of a 20-ms data record. The error bars represent the standard deviation of the amplitude fluctuation within 20 ms. (c) Amplitude ratios (the left y-axis) and calculated error of phase fluctuation (STD of $\Delta\Phi_1$, i.e. $\delta\Delta\Phi_1$) (the right y-axis) for each 20-ms data set. $\Delta\Phi_1$ represents the phase difference between two orthogonally polarized pulses to synthesize the small pulse peaks in (a).

## C. Tuning the polarization modulation scenarios

When the feedback loop is engaged, the phase determining the polarization modulation scenario is locked to the phase ($\Phi_R$) of a reference signal with frequency $f_{rep}/2$ from SG3. Hence, an electrical delay line is inserted after SG3 to tune $\Phi_R$, and thus the polarization modulation scenario. Figure 3 shows the amplitude modulation recorded by DAQ when setting the delay line at 0 ns (Fig. 3(a)) and 2 ns (Fig. 3(b)) respectively. The relative amplitudes of the adjacent pulses change while the amplitude modulation frequency stay the same. The amplitude changes reflect the change of the polarization modulation states. As the polarization modulation frequency is 46.735-MHz, a 2-ns delay corresponds to a phase change of $0.187\pi$. Because of the aforementioned noise floor as a result of artifacts of PD and DAQ detection, it is not feasible to match the polarization tuning exactly with the 2-ns delay. However, the results in Fig. 3 clearly demonstrate that the electrical delay line offers a convenient knob to tune between different polarization modulation scenarios.

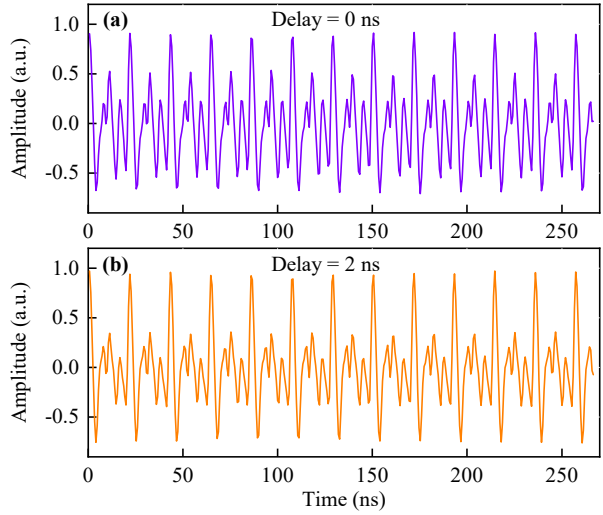

FIG. 3. Polarization-state evolution with different delay times of (a) 0 ns and (b) 2 ns. The amplitude changes between (a) and (b) reflect the change of the polarization modulation states.

## D. Polarization modulation at $N = 4$

The polarization modulation frequency can also be modified by tuning $\Delta f_{ceo}$, e.g. by changing the AOM modulation frequency in the two arms. As an example, a polarization modulation at the frequency of $f_{rep}/4$ is generated. For this purpose, +1st diffraction orders were used for both AOMs, and the driving frequencies were set to be 70 MHz ($f_{AOM1}$) and 93.3675 MHz ($f_{AOM2}$), respectively, which is limited by the modulation bandwidth of the AOMs (85±15 MHz). This parameter configuration provides $\Delta f_{ceo} = f_{AOM2} - f_{AOM1} = 23.3675 MHz = f_{rep}/4$. The result of a phase-stabilized polarization modulation sampled by the same characterization configuration is shown in Fig. 4. Three 190-ns snapshots that are 10-ms apart from each other from the data sequence are presented. The four amplitude peaks within one modulation period denote four different polarization states. The phase of the two adjacent polarization states shifts by $\pi/2$.

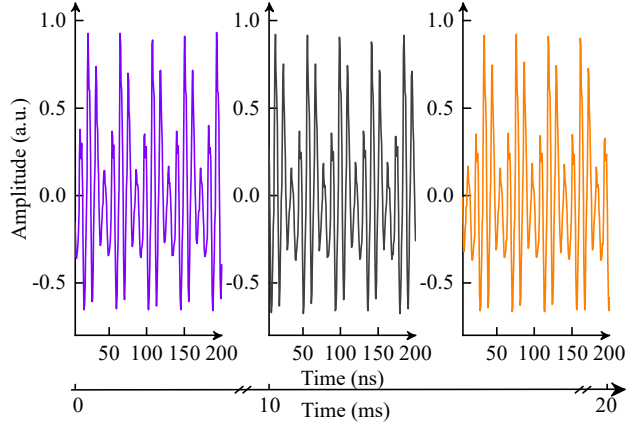

FIG. 4. Project intensity signals for polarization modulation at the frequency of $f_{rep}/4$. Three 190-ns snapshots at around 0 ms, 10 ms and 20 ms from the data sequence are presented, in which $N = 4$ synthesized pulses can be observed in one modulation period.

## IV. DISCUSSION AND CONCLUSIONS

In this paper, we demonstrate a stabilized, high-frequency, polarization-state modulation approach based on optical frequency combs. Two pulse trains split from a $f_{rep}$-stabilized optical frequency comb are used to synthesize the polarization-modulated pulses when their relative phase difference is well controlled. By implementing a feedback loop, the long-term stability of the polarization modulation is demonstrated. With our system, the phase difference fluctuations of six sets of data taken within 3 minutes are all less than 2.6°. An attendant advantage of the electronic feedback loop by imparting a phase correction to $f_{ceo}$ using the AOM, is the wide servo bandwidth, which is currently limited by the PID circuit to roughly 300 kHz. The generated polarization states can also be conveniently adjusted by an electric delay line in the feed-back loop. Moreover, the polarization state modulation frequency can be further modified by changing the driving frequencies of AOMs in the interferometer. A modulation frequency of $f_{rep}/4$ is also demonstrated.

In the demonstrated configuration, the polarization-modulation frequency is determined by $\Delta f_{ceo}$. The $\Delta\Phi$ between adjacent synthesized pulses evolves as $2\pi\Delta f_{ceo}/f_{rep}$. Thus, there is a trade-off between the fast polarization modulation and the number of discrete polarization states generated for a fixed repetition rate. To achieve faster modulation between a greater number of discrete states, frequency combs with high repetition rates need to be used, such as microresonator combs or semiconductor-based combs with repetition rates from 10s of GHz into the THz.[45-47] Furthermore, not only can we tune the frequency at which the polarization modulation occurs, but we can also tune the polarization states by interfering the two orthogonally polarized pulses. This ability allows for the extraction of the exact response of a sample to each separate polarization state by relating the phase and amplitude of the optical or electronic response of the sample to the applied phase of the polarization modulation. With very high modulation frequency, such measurements are sensitive to both polarization and response speed. In the future, this technique could be combined with other multi-comb spectroscopies[48,49] to multiplex polarization dependent studies in a wide array of physical systems. The simplicity of the technique, together with the precise control over optical spectra that frequency combs enable, promises versatile opportunities along this direction.

Rapid polarization modulation at RF frequencies can also be achieved by using an electro-optic modulator (EOM). Nevertheless, the EOM approach has at least the following two drawbacks comparing to the coherent synthesis approach demonstrated in this work: firstly, unless the modulation frequency is tuned to match half of $f_{rep}$, the phase difference between the successive pulses varies, when a sinusoidal driving signal is applied to the EO crystal. By contrast, the phase difference varies uniformly in the coherent synthesis approach based on OFCs; secondly, the optical bandwidth of EOM is very limited due to the dispersion of the EO crystal, which cause phase retardation for wavelength off the designed center. However, the similar dispersion issue of the AOMs used in the scheme of this work can be easily eliminated using a double-pass configuration.

We emphasize that the stabilized high-frequency polarization modulation is a powerful optical source that is widely applicable to study the polarization-sensitive response of interesting quantum materials. Many topological semimetals, e.g. $TaIrTe_4$ and $Mo(W)Te_2$[13,14] and 2D layered semiconductors, e.g. black phosphorus and $ReS_2$[15,16] are highly anisotropic and thus sensitive to the direction of linear polarization. On the other hand, many novel quantum degrees of freedom of quantum materials, such as spin, valley and chirality, are sensitive to the helicity of light.[7,11,12] The capability of radio frequency polarization modulation, especially circular polarized modulation, could not only enable sensitive characterization and thus advance the understanding of fundamental quantum physics, but also provide rapid optical manipulation of these interesting physics properties of quantum materials.

## ACKNOWLEDGMENTS

R. Zhang would like to acknowledge the support from China Scholarship Council. D. Sun would like to acknowledge the support from Beijing Natural Science Foundation (Grant Nos: JQ19001).

**DATA AVAILABILITY:** The data that support the findings of this study are available from the corresponding authors upon reasonable request.